%% file: document.tex
\documentclass[aps,pra,twocolumn,10pt,superscriptaddress]{revtex4-1} 
\pdfoutput=1
\usepackage{amsmath}
\usepackage{amsfonts}
\usepackage{amssymb}
\usepackage{graphicx}
\usepackage{microtype}
\usepackage{booktabs}
\usepackage[latin1]{inputenc}

\usepackage[svgnames]{xcolor}

\usepackage[pdftex]{hyperref}
\hypersetup{colorlinks=true, linkcolor=blue, citecolor=ForestGreen, urlcolor=DarkOrchid}

\usepackage{xcolor}
\usepackage{graphicx}

\usepackage{amsmath, amssymb, amsfonts}
	
	\newcommand{\R}{\mathbb{R}}

\usepackage{mathtools}
	\mathtoolsset{showonlyrefs=true}				

\usepackage{chngcntr}
	\counterwithout{equation}{section}				

\begin{document}

%
\title{Flight Gate Assignment with a Quantum Annealer
}


\author{Tobias Stollenwerk}
\affiliation {German Aerospace Center (DLR), Linder Höhe, 51147 Köln, Germany}
\author{Elisabeth Lobe}
\affiliation {German Aerospace Center (DLR), Linder Höhe, 51147 Köln, Germany}
\author{Martin Jung}
\affiliation {German Aerospace Center (DLR), Linder Höhe, 51147 Köln, Germany}



\begin{abstract}
	\input{abstract}
\end{abstract}
\maketitle              

%
\input{content}

%
\bibliography{references}

\end{document}

%% file: abstract.tex

Optimal flight gate assignment is a highly relevant optimization problem from airport management.
Among others, an important goal is the minimization of the total transit time of the passengers.
The corresponding objective function is quadratic in the binary decision variables encoding the flight-to-gate assignment.
Hence, it is a quadratic assignment problem being hard to solve in general.
In this work we investigate the solvability of this problem with a D-Wave quantum annealer.
These machines are optimizers for quadratic unconstrained optimization problems (QUBO).
Therefore the flight gate assignment problem seems to be well suited for these machines.
We use real world data from a mid-sized German airport as well as simulation based data to extract typical instances small enough to be amenable to the D-Wave machine.
In order to mitigate precision problems, we employ bin packing on the passenger numbers to reduce the precision requirements of the extracted instances.
We find that, for the instances we investigated, the bin packing has little effect on the solution quality.
Hence, we were able to solve small problem instances extracted from real data with the D-Wave 2000Q quantum annealer.

%% file: content.tex

\section{Introduction}

Modern airport management requires a more holistic approach to the whole travel chain, aiming at a better situational awareness of airport landside processes and an improved resource management. 
One main key to achieve this is proactive passenger management \cite{classen2015}.
Unlike common reactive approaches, proactive passenger management utilizes an early knowledge about the passengers' status and the expected situation in the terminal along with resulting system loads and resource deployment, e.g. using simulation techniques like in \cite{jung2017}. 
An appropriate and modern management also considers dependencies of airside and landside operations as well as costs and performance. 
One part of the proactive passenger management is a proper flight gate assignment. 
Especially large connecting hub airports have to deal with the still growing demand of traffic and rising numbers of passengers and baggage that needs to be transferred between flights. 
Although there are several objective functions in the field of airport planning, we focus on reducing the total transit time for passengers in an airport \cite{kim2017}.
The main goals are the increase of passenger comfort and punctuality. 
This problem is related to the quadratic assignment problem, a fundamental problem in combinatorial optimization whose standard formulation is NP-hard \cite{garey2002}. 

In general, most planning problems belong to the class of discrete optimization problems which are hard to solve classically.
Therefore it is worth studying new hardware architectures, like quantum computers, which may outperform classical devices.
The first commercially available quantum annealer, developed by the Canadian company D-Wave Systems, is a heuristic solver using quantum annealing for optimizing quadratic functions over binary variables without further constraints (QUBOs).
A wide range of combinatorial optimization problems can be brought into such a QUBO format by standard transformations \cite{hammer2012}.
But usually these transformations produce overhead, e.g. by increasing the number of variables, the required connections between them or the value of the coefficients. 
It is important to assess the impact of this overhead in order to estimate the performance of future devices \cite{rieffel2015}.
In this work, we investigate the solvability of the optimal flight gate assignment problem with a D-Wave 2000Q quantum annealer.

The paper is structured as follows:
In section \ref{sec:problem_definition} we formally introduce the problem preparing the mapping to QUBO, which is done in section \ref{sec:qubo_mapping}.
The extraction of smaller, but representative problem instances is covered in section \ref{sec:instance_extraction}. 
In section \ref{sec:quantum_annealing} we present our results for solving these smaller problem instances with a D-Wave 2000Q quantum annealer.

\section{Formal Problem Definition}		
\label{sec:problem_definition}

Flight gate assignment was already addressed in different versions \cite{haghani1998}\cite{mangoubi1985}.
The following mainly corresponds to the formulation from \cite{kim2017} as a quadratic binary program with linear constraints. 


\subsection{Input Parameters}

The typical passenger flow in an airport can usually be divided into three parts: 
After the airplane has arrived at the gate, one part of the passengers passes the baggage claim and leaves the airport. 
Other passengers stay in the airport to take connecting flights. 
These transit passengers can take up a significant fraction of the total passenger amount.
The third part are passengers which enter the airport through the security checkpoint and leave with a flight.  
The parameters of the problem are summarized in table~\ref{table:input}.  

\newcommand{\dep}{^\text{dep}}
\newcommand{\arr}{^\text{arr}}
\newcommand{\arrdep}{^\text{arr/dep}}
\newcommand{\iin}{^\text{in}}
\newcommand{\out}{^\text{out}}
\newcommand{\buf}{^\text{buf}}
\newcommand{\one}{^\text{one}}

\begin{table}[bt] 
\centering
\caption{Input data for a flight gate assignment instance}\label{table:input}
\begin{tabular}{|l|l|}\hline
	$F$					& Set of flights ($i \in F$)\\\hline
	$G$					& Set of gates ($\alpha \in G$)\\\hline
    $n\dep_i$    		& No. of passengers departing with flight $i$ \\\hline
    $n\arr_i$      		& No. of passengers arriving with flight $i$ \\\hline
    $n_{ij}$     		& No of transfer passengers between flights $i$ and $j$ \\\hline
    $t\arr_\alpha$ 		& Transfer time from gate $\alpha$ to baggage claim \\\hline
    $t\dep_\alpha$ 		& Transfer time from check-in to gate $\alpha$ \\\hline
    $t_{\alpha\beta}$ 	& Transfer time from gate $\alpha$ to gate $\beta$ \\\hline
    $t\iin_i$ 			& Arrival time of flight $i$ \\\hline
    $t\out_i$ 			& Departure time of flight $i$ \\\hline
    $t\buf$			 	& Buffer time between two flights at the same gate \\\hline
\end{tabular}
\end{table}

\subsection{Variables and Objective}

Assignment problems can easily be represented with binary variables indicating whether or not a resource is assigned to a certain facility.
The variables form a matrix indexed over the resources and the facilities. 
The binary decision variables are $x \in \{0,1\}^{F \times G}$ with
\begin{equation}
	x_{i \alpha} = \begin{cases}
        1, &\text{if flight $i \in F$ is assigned to gate $\alpha \in G$}, \\
						0, &\text{otherwise}.
					\end{cases}
\end{equation}

Like already stated, the passenger flow divides into three parts and so does the objective function: 
The partial sums of the arriving, the departing and the transfer passengers sum up to the total transfer time of all passengers. 
For the arrival part we get a contribution of the corresponding time $t\arr_\alpha$ for each of the $n\arr_i$ passengers if flight $i$ is assigned to gate $\alpha$.
Together with the departure part, which is obtained analogously, the linear terms of the objective are
\begin{equation}
	T\arrdep(x) = \sum_{i\alpha} n\arrdep_i t\arrdep_\alpha \, x_{i\alpha}.
\end{equation}
The contribution of the transfer passengers is the reason for the hardness of the problem: 
Only if flight $i$ is assigned to gate $\alpha$ and flight $j$ to gate $\beta$ the corresponding time is added.  
This results in the quadratic term
\begin{equation}
 	T^\text{trans}(x) = \sum_{i\alpha j\beta} n_{ij} t_{\alpha\beta} \, x_{i\alpha}x_{j\beta}.
\end{equation}
The total objective function is
\begin{equation} \label{eq:objective}
	T(x) = T\arr(x) + T\dep(x) + T^\text{trans}(x).
\end{equation}	

\subsection{Constraints}\label{sec:constraints}

Not all binary encodings for $x$ form valid solutions to the problem.
There are several further restrictions which need to be added as constraints.  
In this model a flight corresponds to a single airplane arriving and departing at a single gate. 
It is obvious, that every flight can only be assigned to a single gate, therefore we have
\begin{equation}
    \sum_\alpha x_{i\alpha} = 1 \qquad\forall i \in F.\label{const:one}
\end{equation}

Furthermore it is clear that no flight can be assigned to a gate which is already occupied by another flight at the same time. 
These forbidden flight pairs with overlapping time slots can be aggregated in 
\begin{equation}
    P = \left\{(i, j) \in F^2 : t^\text{in}_i < t^\text{in}_j < t^\text{out}_i + t^\text{buf} \right\}.
\end{equation}
The resulting linear inequalities $x_{i\alpha} + x_{j\alpha} \leq 1$ are equivalent to the quadratic constraints
\begin{equation}
            x_{i\alpha} \cdot x_{j\alpha} = 0 \qquad\forall (i, j) \in P ~\forall \alpha \in G .\label{const:not}
\end{equation}

\section{Mapping to QUBO}
\label{sec:qubo_mapping}

QUBOs are a special case of integer programs minimizing a quadratic objective function over binary variables $x \in \{0,1\}^n$. 
The standard format is the following
\begin{equation}\label{eq:qubo}
	q(x) = x^\top Q x = \sum_{j=1}^n Q_{jj} x_j + \sum_{\substack{j,k = 1 \\ j<k}}^n Q_{jk} x_j x_k 
\end{equation}
with an upper-triangular quadratic matrix $Q \in \mathbb{R}^{n \times n}$.
While the presented objective function already follows this format the constraints need to be reduced which is shown in this section.

\subsection{Penalty Terms}

The standard way to reduce constraints, like already shown in e.g. \cite{rieffel2015}, is to introduce terms penalizing those variable choices that violate the constraints. 
Just in these cases a certain positive value is added to the objective function to favor valid configurations while minimizing. 
The quadratic terms 
\begin{equation} 
	C\one(x) = \sum_i \bigg( \sum_\alpha x_{i\alpha} - 1 \bigg)^2
\end{equation}
and
\begin{equation} 
    C^\text{not}(x) = \sum_\alpha \sum_{(i, j) \in P} x_{i\alpha} x_{j\alpha} 
\end{equation}
fulfill
\begin{equation}
	C^{\text{one}/\text{not}}
			\begin{cases}
				> 0, &\text{if constraint is violated,}\\
				= 0, &\text{if constraint is fulfilled,} 
			\end{cases}
\end{equation}
and therefore are suitable penalty terms which can be combined with the objective function. 
Since the benefit in the objective function in case of an invalid variable choice should not exceed the penalty, two parameters $\lambda_\text{one}, \lambda_\text{not} \in \R_+$ need to be introduced:   
\begin{equation}
	q(x) = T(x) + \lambda\one C\one(x) + \lambda^\text{not} C^\text{not}(x). 
\end{equation}
In theory these parameters could be set to infinity, but in practice this is not possible and they have to be chosen carefully.  

\subsection{Choosing Penalty Weights}

The parameters $\lambda\one$ and $\lambda^\text{not}$ need to be large enough to ensure that a solution always fulfills the constraints.
However due to precision restrictions of the D-Wave machine it is favorable to choose them as small as possible.
In this section we present two different possibilities to obtain suitable values. 
Besides a worst case analysis for each single constraint a bisection algorithm which iteratively checks penalty weights against the solution validity is presented.
In the following we call a variable to be activated if it is set to one.    

\subsubsection{Worst Case Estimation}

The minimal contribution of $C\one$ and $C^\text{not}$ when breaking the corresponding constraint is just one. 
Therefore the corresponding minimal penalties are $\lambda\one$ and $\lambda^\text{not}$.
Since the objective function and $C^\text{not}$ only contain non-negative coefficients, just $C\one$ enforces some of the variables to be set to one.
Therefore activating more than one $x_{i\alpha}$, for one flight $i$, does not improve the objective value. 
Hence $\lambda\one$ just needs to exceed the benefit of deactivating a single variable in a valid solution. 
This variable usually appears in several summands of the objective function. 
Hence in the worst case the objective could be reduced by the sum of all coefficients of monomials including this variable, which is 
\begin{equation}
	T_{i\alpha}(x) \vcentcolon = n\dep_i t\dep_\alpha + n\arr_i t\arr_\alpha + \sum_{j} n_{ij} \sum_{\beta}t_{\alpha\beta} x_{j\beta}.
\end{equation}
Assuming the penalty is chosen large enough it will also be sufficient for all gates $\beta$ appearing in $T_{i\alpha}$.
Therefore in the last part of the sum, for every flight $j$, just one gate $\beta$ needs to be taken into account for which $x_{j\beta}$ is one.
In the worst case this is the one with the maximal time $t_{\alpha\beta}$, which results in 
\begin{equation}
	T\one \vcentcolon = \max_{i, \alpha} \Big(n\dep_i t\dep_\alpha + n\arr_i t\arr_\alpha + \max_{\beta} t_{\alpha\beta} \sum_{j} n_{ij} \Big).
\end{equation}

Considering $C^\text{not}$, it is not preferable to add an additional flight pair that is forbidden. 
But if the penalty is not large enough, assigning one flight $i$ to a gate $\gamma$ although this gate is already occupied by another flight $j$ rather than using the different gate $\alpha$ could reduce the objective value.
This means $x_{i\gamma} = 1$ might be preferred to $x_{i\alpha} = 1$ although $(i,j) \in P$ and $x_{j\gamma} = 1$.
The resulting benefit can be calculated from the difference of $T_{i\alpha}$ and $T_{i\gamma}$.
For the estimation of the worst case this can be simplified by taking the maximum transfer time for $T_{i\alpha}$ and the minimum for $T_{i\gamma}$: 
\begin{equation}
\mathclap{
\begin{aligned}
	T^{\text{not}} \vcentcolon = \max_{i,\alpha,\gamma} \bigg(\begin{alignedat}[t]{2}
																	\!&\Big(n\dep_i t\dep_\alpha + n\arr_i t\arr_\alpha + \max_{\beta} t_{\alpha\beta} \sum_{j} n_{ij} \Big) \\ 
															  		\!-& \Big(n\dep_i t\dep_\gamma + n\arr_i t\arr_\gamma + \min_{\beta} t_{\gamma\beta} \sum_{j} n_{ij} \Big)\!\bigg)
															  \end{alignedat} \\
					= \max_{i,\alpha,\gamma}\bigg(
                    \!\big(n_i^\text{dep} t^\text{dep}_\alpha - n_i^\text{dep} t^\text{dep}_\gamma \big) 
                                                        +\big(n_i^\text{arr} t^\text{arr}_\alpha - n_i^\text{arr} t^\text{arr}_\gamma\big) \\
                                                        \!+\max_{\beta}\big(t_{\alpha \beta}-t_{\gamma\beta}\big)\sum_{j}n_{ij}\!
                                                        \bigg)\!.
\end{aligned}}
\end{equation}

All in all using $\lambda\one = T\one + \varepsilon$ and $\lambda^\text{not} = T^\text{not} + \varepsilon$ for some $\varepsilon > 0$ ensures that the minimum of $q$ satisfies the constraints. 
The provided boundaries can be calculated easily, but unfortunately they may take pretty large values depending on the given parameters.
Since usually the worst case is also not the most probable it is a very rough estimation and there is some room for improvement.

\subsubsection{Bisection Method}
\label{sec:bisection}

If the constraints are independent of each other, a simple bisection method can be used to find an approximation of the boundary between valid and invalid penalty values. 
In the course of this, all but one penalty is fixed, while the bisection is employed in one dimension at a time.
These fixed penalty values as well as an upper starting point of the bisection method need to yield valid solutions.
For given values of the penalties, a classical solver like SCIP is used to find the exact solution \cite{gleixner2017}.
But it is imaginable that also the D-Wave machine itself could be used. 
For very small instances it might happen that one of the constraints becomes redundant and therefore is always fulfilled if the other one is.
In these cases just the remaining constraint needs to be evaluated.   
This method, may lead to much smaller values than the worst case estimation.
However it requires solving the problem several times exactly for each instance.
Therefore this approach is not viable in an operational setting.

\section{Instance Extraction and Generation}
\label{sec:instance_extraction}

\begin{figure}[htb]
    \centering
    \includegraphics[width=0.9\linewidth]{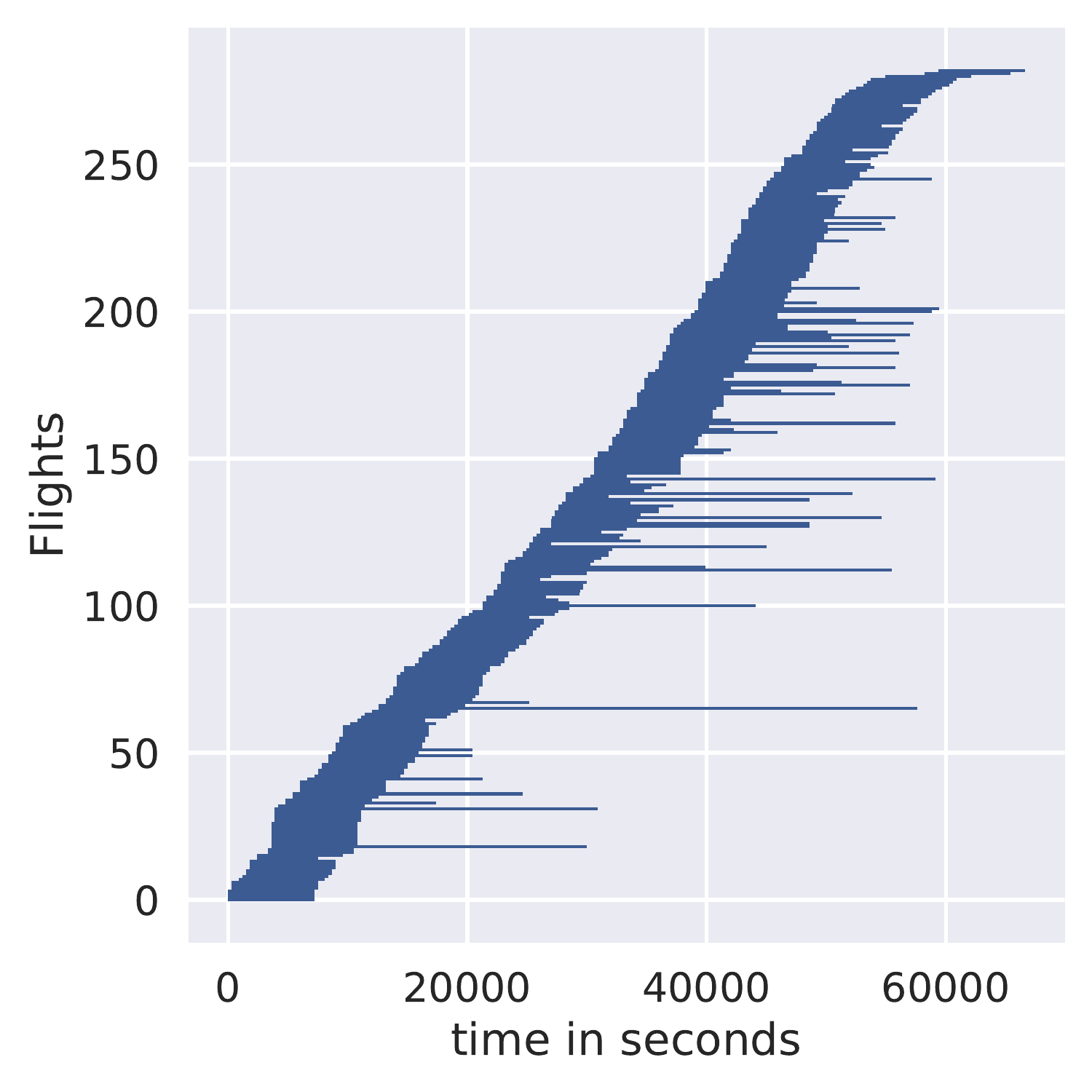}
    \includegraphics[width=0.8\linewidth]{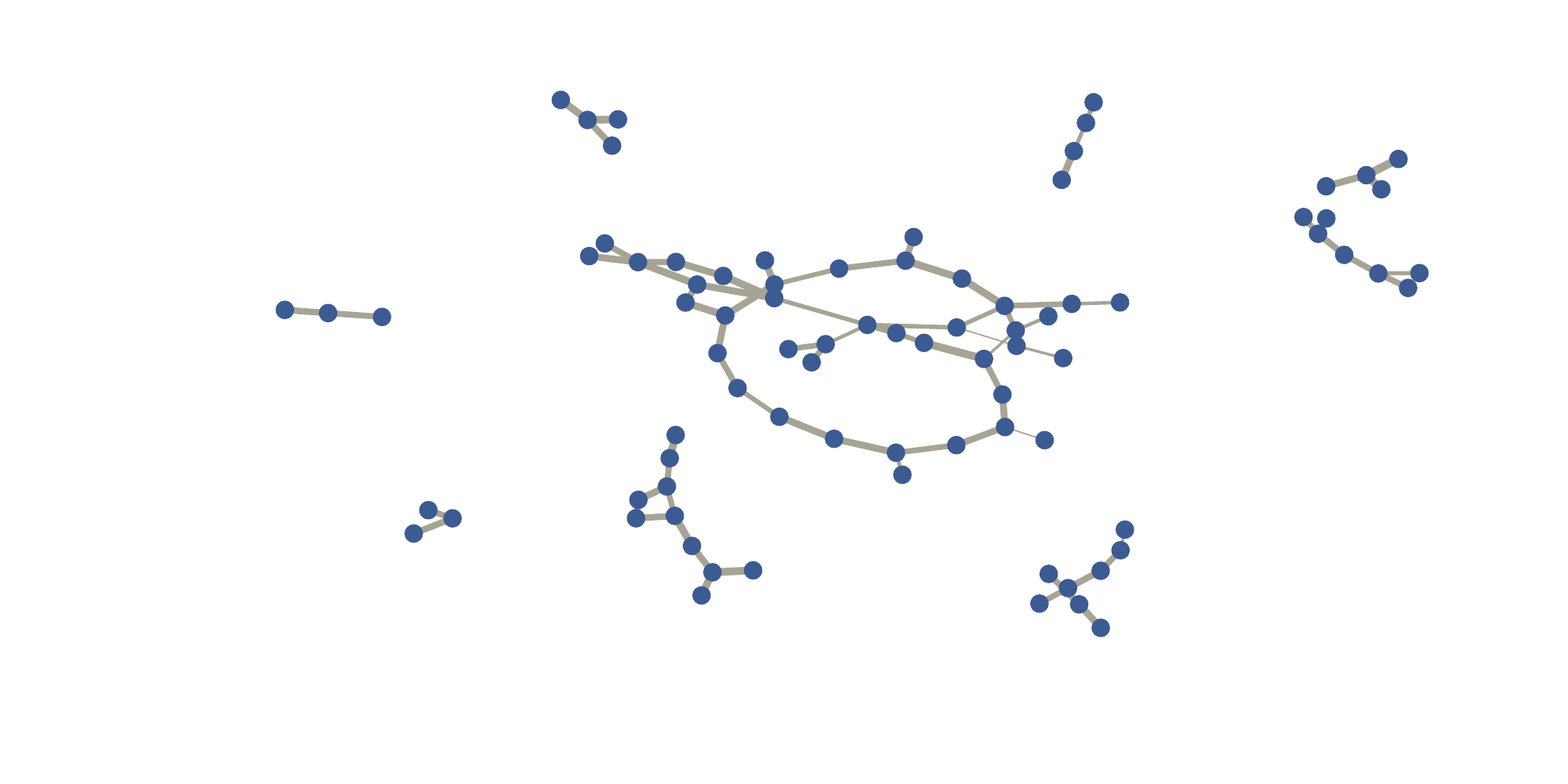}
    \caption{Top: Time at the airport for all flights during a whole day on a mid-sized German airport. 
    Bottom: Graph representing the transfer passengers for a whole day. Each vertex is a flight, each edge are transfer passengers between two flights. }
    \label{fig:main_problem_instance}
\end{figure}

From \cite{jung2017}, applying agent based simulation techniques of \cite{bonabeau2002}, we extract the data to estimate the transfer times 
$t\buf$, $t\dep_\alpha$, $t\arr_\alpha$ $\forall \alpha\in G$ and $t_{\alpha\beta}$ $\forall (\alpha, \beta) \in G^2$.
The second part of our data source consists of a flight schedule of a mid-sized German airport for a full day.
This gives us the number of passengers $n\dep_i$, $n\arr_i$ $\forall i\in F$ and $n_{ij}$ $\forall (i, j) \in F^2$. 
The resulting problem instance is depicted in figure~\ref{fig:main_problem_instance}.

\subsection{Preprocessing}

Since the given data set contains a slice of a flight schedule cut off at one day, there are flights which either do not have an arrival or a departure time. 
These flights are removed.
Furthermore for some of the airplanes, the standing time is so long that it is reasonable to assume the airplane is moved away from the gate for some time before returning to a possibly different gate.
Hence, we considered all flights with a time at the airport above $120$ minutes as two separate flights.
In order to extract typical, hard instances from the data, we removed all flights with no transfer passengers.
The remaining instance now consists of 89 flights with 80 transfers and over 35 gates. 
However, since this corresponds to more than 3000 binary variables further subdivision is needed to make the problem amenable to the current D-Wave 2000Q hardware.

A way to achieve this becomes apparent when visualizing the dependencies among flights with transfer passengers in a graph as in the bottom part of figure~\ref{fig:main_problem_instance}.
It can be observed that the graph is divided into several connected components, various smaller ones with 3 to 11 flights and a single larger one.
The time intervals are much more distributed.   
Therefore extracting these special subgraphs provides suitable instances to be tested. 
In addition to using the connected components of the transfer passenger graph, we randomly cut the largest connected component to create larger test instances.
This results in $163$ instances with number of flights and gates from $3$ to $16$ and $2$ to $16$, respectively.
The set of these instances will be denoted by $\mathcal{I}_\text{CC}$.

An alternative to using only flights with transfer passengers, is to use only flights within a certain time interval.
However, this option has two major disadvantages: 
In such a time slot most of the flight intervals overlap mutually and therefore each flight requires a different gate. 
Also, in our data set, these flights do have almost no transferring passengers. 
Both issues simplify the subproblems severely.
Therefore this alternative is not pursued in this work.
		
\subsection{Bin Packing}
As we will see, some of the instances from $\mathcal{I}_\text{CC}$ are small enough to fit on the D-Wave 2000Q machine.
However, they exhibit a strong spread of the coefficients in the QUBO.
It is known that this can suppress the success probability due to the limited precision of the D-Wave machine \cite{stollenwerk2017}.
Therefore, we tried to reduce the spread of the coefficients while retaining the heart of the problem as best as possible.
First, we mapped the passenger numbers to natural numbers in $\{0, 1, \dots, N_\text{p}\}$ with bin packing, where $N_\text{p}$ is the number of bins. 
Moreover we mapped the transfer times to random natural numbers from $\{1, \dots, N_\text{t}\}$.
This is reasonable, since the original time data was drawn from a simulation data, which showed similar behavior.
The mapping of the maximum transfer time in the instance to $N_\text{t}$ introduces a scaling factor to the objective function, which is irrelevant for the solution.
In order to assess the impact of the bin packing for the number of passengers, we solved each problem before and after bin packing with the exact solver SCIP.
Let the solutions before and after bin packing be denoted as $\mathbf{x}$ and $\mathbf{\hat{x}}$.
The approximation ratio is then given by the $R=\frac{T(\mathbf{\hat{x}})}{T(\mathbf{x})}$. 
Where $T$ is the objective function \eqref{eq:objective} before bin packing.
If $R=1$, the bin packing has no effect on the solution quality.
\begin{figure}[hbt]
    \centering
    \includegraphics[width=0.9\linewidth]{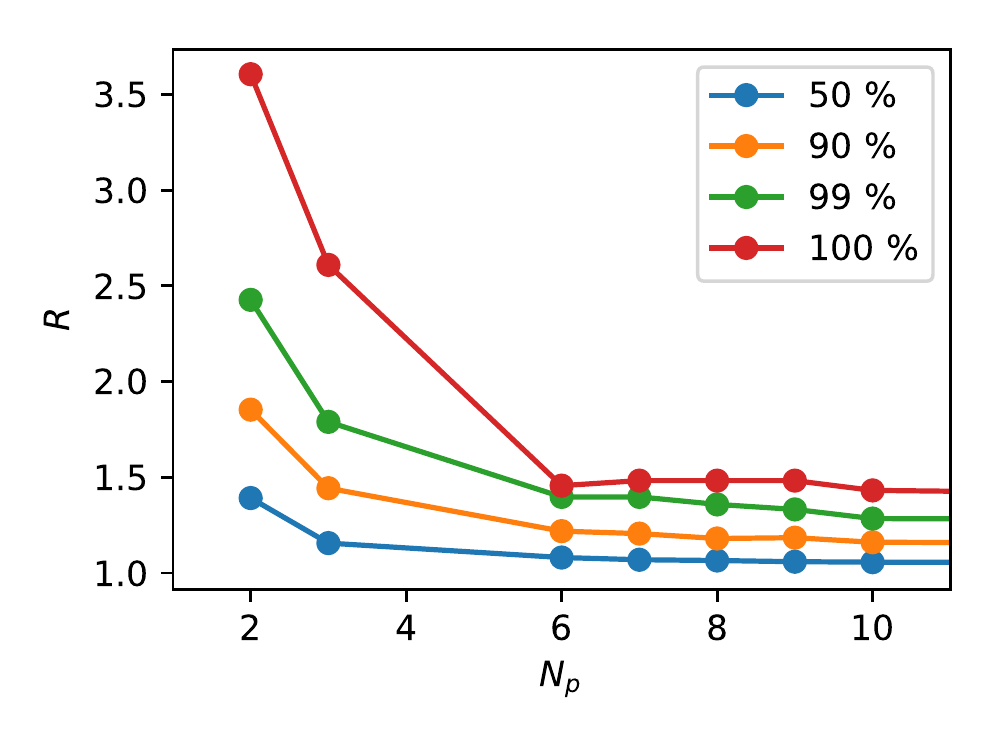}
    \includegraphics[width=0.9\linewidth]{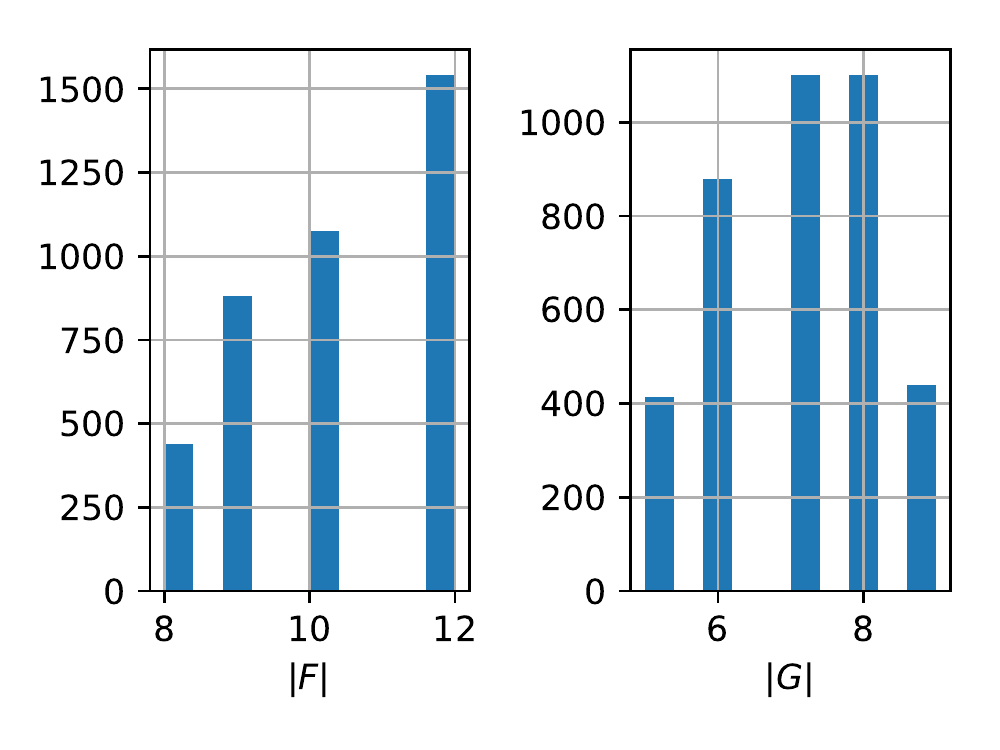}
    \caption{Top: Approximation ratio for bin packed instances.
    The different curves show percentiles.
    Bottom: Distribution of number of flights and gates $|F|$, $|G|$.}
    \label{fig:bin_packing_impact}
\end{figure}
Figure~\ref{fig:bin_packing_impact} shows the approximation ratio for various bin packed instances.
We used values of $N_\text{p} \in \{2, 3, 6, 7, 8, 9, 10\}$ and $N_\text{t} \in \{2, 3, 6, 10\}$ and all combinations thereof.
Moreover, we restrict ourselves to instances with $|F|\cdot|G|<100$.
One can see, that the approximation ratio is close to one for the majority of the bin packed instances. 
Meaning, the bin packing has little effect on the solution quality, at least for the instances we investigated.

For the investigation of quantum annealing, we restrict ourselves to $N_\text{p}, N_\text{t} \in \{2, 3, 6, 10\}$.
The corresponding instances will be denoted by $\mathcal{I}_\text{BP}$.

\section{Quantum Annealing}
\label{sec:quantum_annealing}
\subsection{Embedding}
\label{sub:sec_embedding}
\begin{figure}[b]
    \centering
    \includegraphics[width=0.9\linewidth]{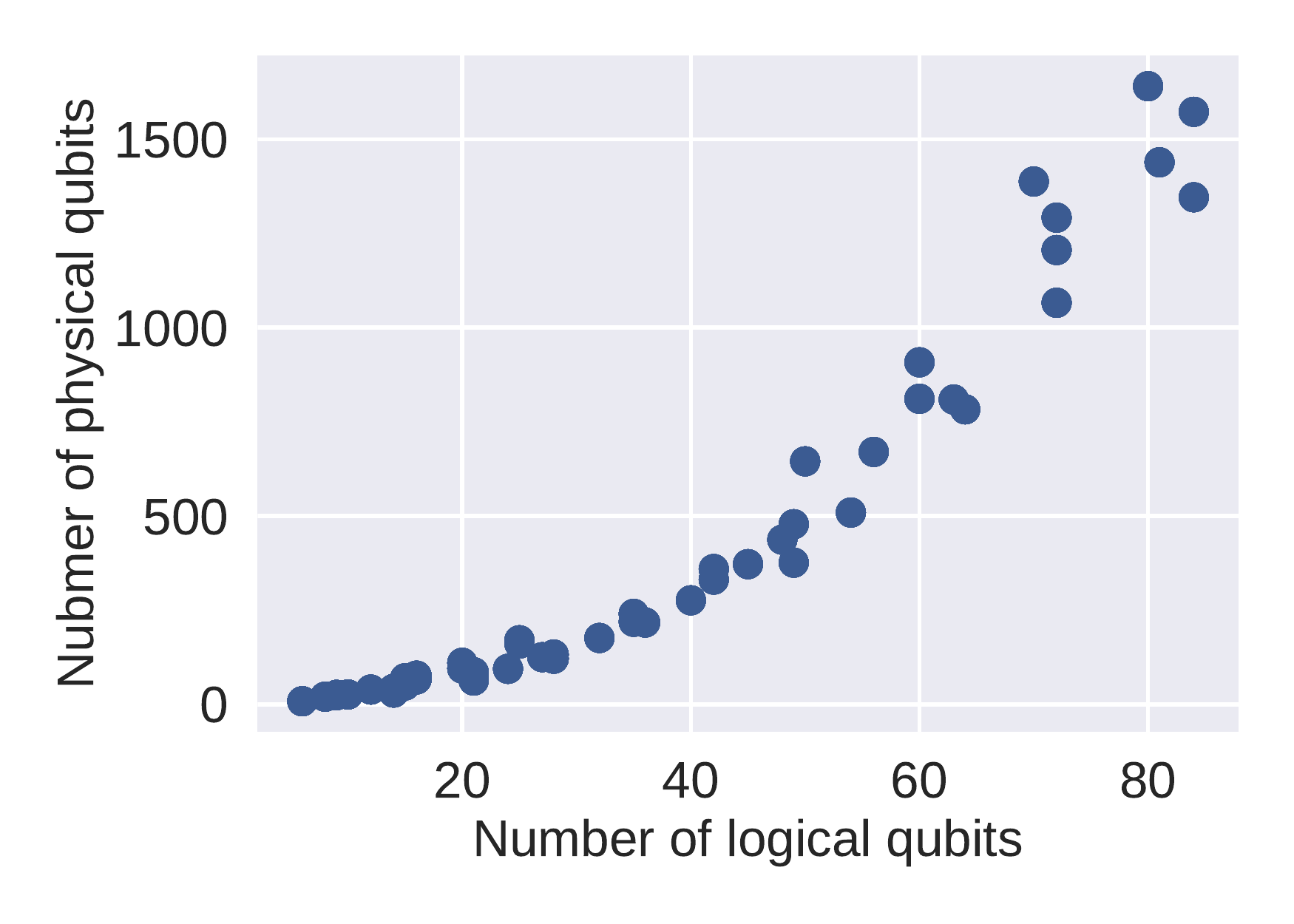}
    \includegraphics[width=0.9\linewidth]{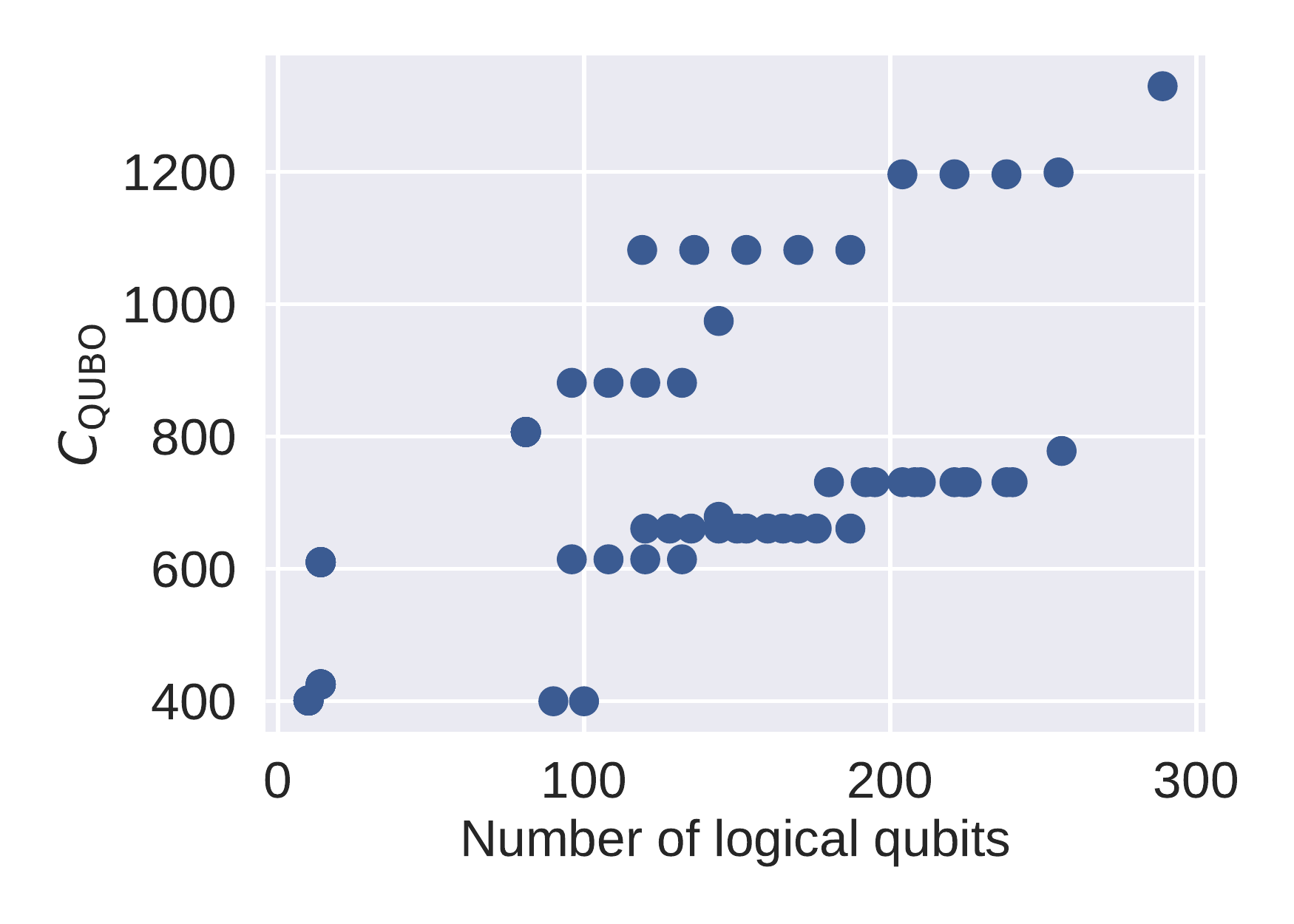}
    \caption{Top: Number of logical and physical qubits on D-Wave 2000Q for instances $\mathcal{I}_\text{CC}$.
    Bottom: Maximum coefficient ratio of the QUBOs for instances $\mathcal{I}_\text{CC}$.}
    \label{fig:embedding_precision}
\end{figure}

The hardware layout of the D-Wave 2000Q quantum annealer restricts the connections between binary variables to the so called Chimera graph \cite{mcgeoch2013}.
In order to make problems with higher connectivity amenable to the machine, we employ minor-embedding \cite{cai2014}.
This includes coupling various physical qubits together into one logical qubit, representing one binary variable, with a strong ferromagnetic coupling $J_F$ in order to ensure that all physical qubits have the same value after readout \cite{Venturelli2015}.
Since the constraint~\eqref{const:one} introduces $|F|$ complete graphs of size $|G|$, we expect at most quadratic increase in the number of physical qubits with the number of logical qubits, which is $|F|\cdot|G|$. 
This is supported by our findings in the top part of figure~\ref{fig:embedding_precision}. 
We calculated five different embeddings for each problem instance from $\mathcal{I}_\text{CC}$.
With this, we were able to embed instances up to $84$ binary variables.

\subsection{Precision Requirements}
\label{sub:sec_precision}
The D-Wave 2000Q has a limited precision in specifying the linear and quadratic coefficients of the Ising model.
As it was shown in \cite{stollenwerk2017}, this can be an inhibiting factor for solving some problems on the D-Wave 2000Q.
In order to assess the precision requirements for each instance, we introduce the maximum coefficient ratio for a QUBO like \eqref{eq:qubo} as
\begin{equation*}
    C_\text{QUBO} = \frac{\max_{ij} | Q_{ij} |}{\min_{ij} | Q_{ij} |} \, ,
\end{equation*}
and for the corresponding embedded Ising model $\sum_i h_i s_i + \sum_{ij} J_{ij} s_i s_j$  as
\begin{equation*}
    C_\text{Ising} = \max\left\{\frac{\max_i | h_i |}{\min_i | h_i |} ,\frac{\max_{ij} | J_{ij} |}{\min_{ij} | J_{ij} |} \right\} \, .
\end{equation*}
The bottom part of figure~\ref{fig:embedding_precision} shows the maximum coefficient ratio for all QUBO from $\mathcal{I}_\text{CC}$. 
We used minimal sufficient penalty weights calculated by bisection as it was described in section \ref{sec:bisection}. 
The corresponding values for the embedded Ising model are orders of magnitude larger (not shown).
Therefore the success probability for these instances is highly suppressed.
However, this does not seem to be sufficient to reduce the precision requirements of the instances to an acceptable level.
In order to mitigate the problem, we will use the bin packed instances $\mathcal{I}_\text{BP}$ for the remainder of this work.

\subsection{Quantum Annealing Results}

\begin{figure}[htb]
    \centering
    \includegraphics[width=0.9\linewidth]{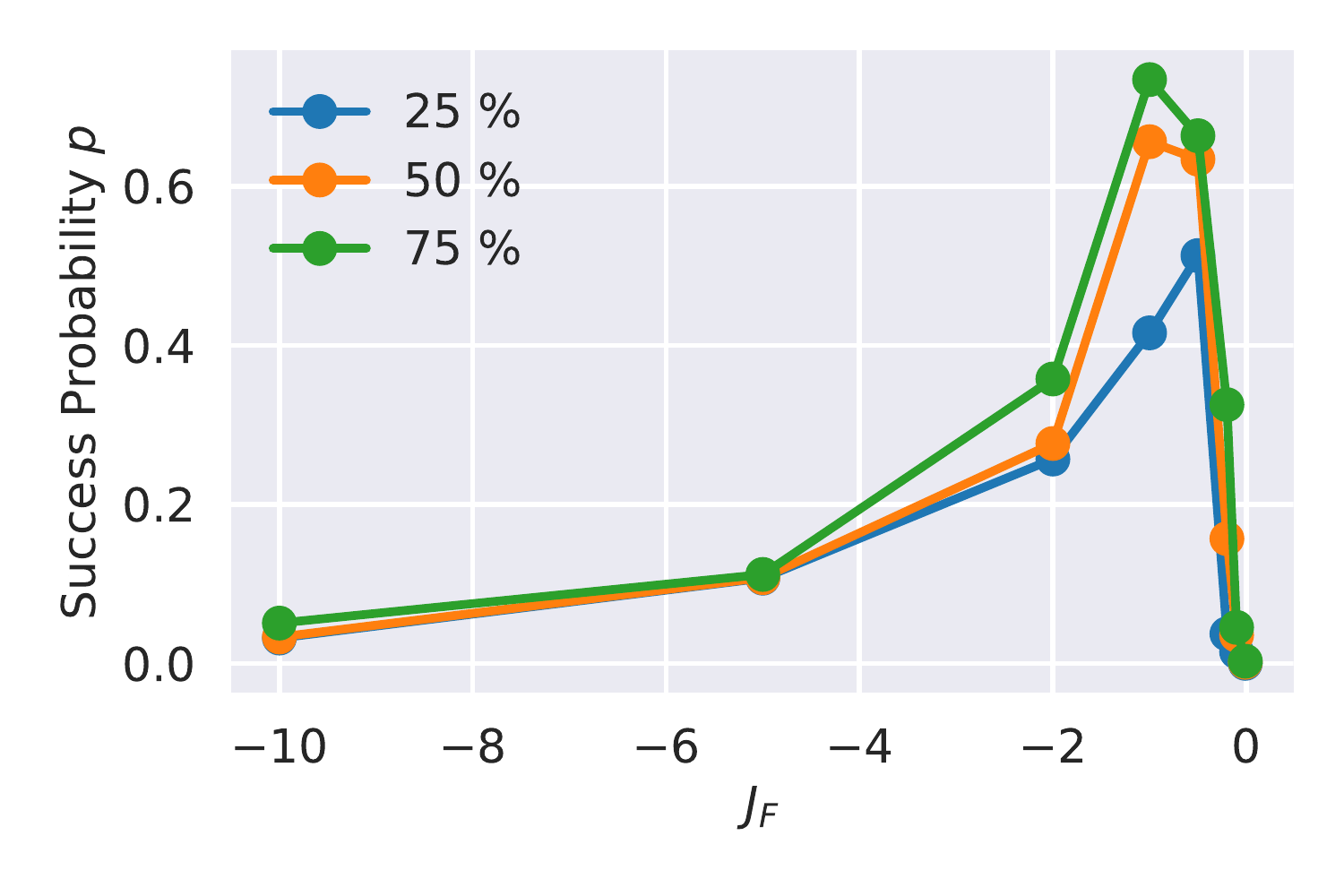}
    \includegraphics[width=0.9\linewidth]{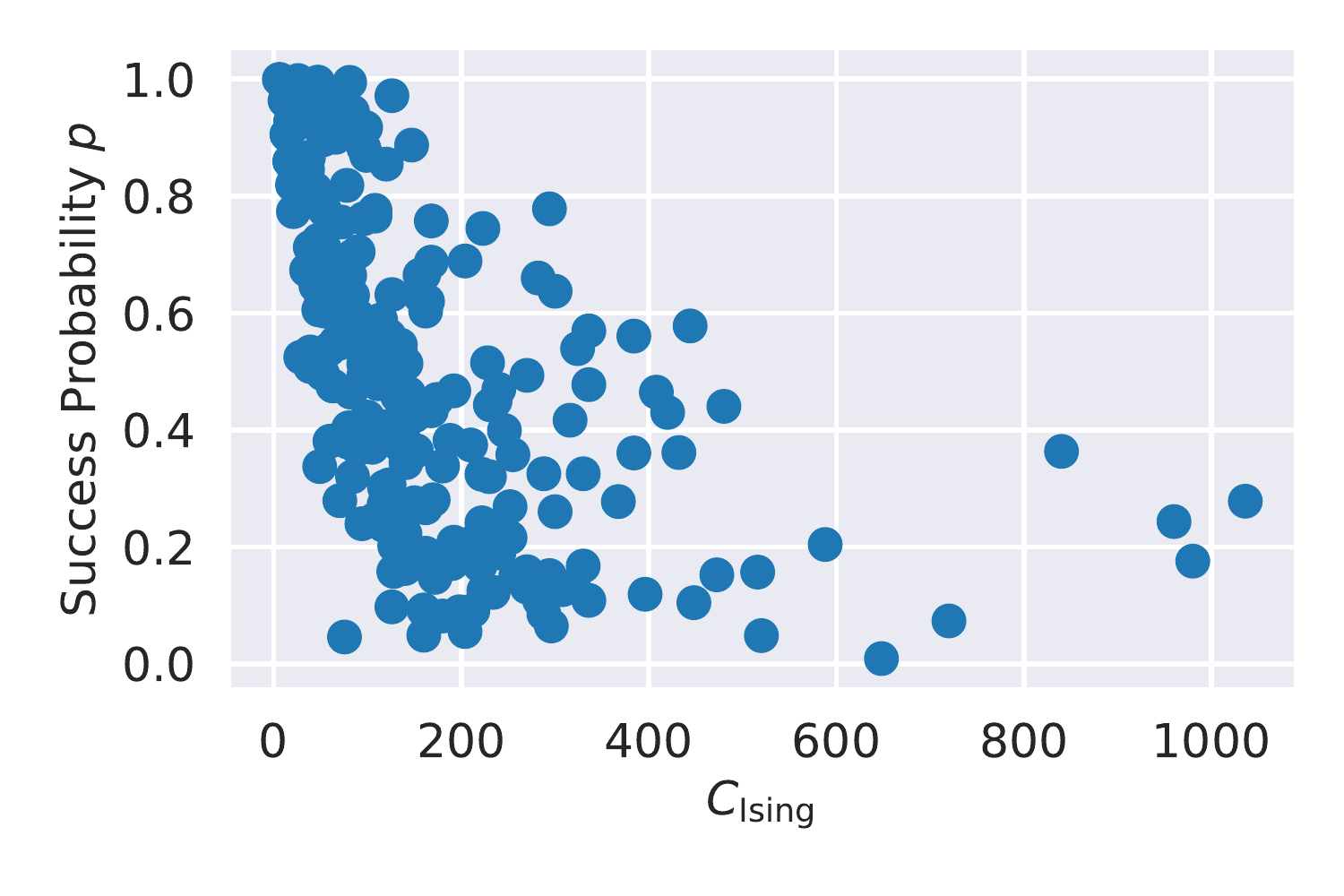}
    \caption{Top: Success probability against the intra-logical qubit coupling in units of the largest coefficient of the embedded Ising model. The data is for a representative instance from  $\mathcal{I}_\text{BP}$.
    The different colors represent the $25$th, $50$th and $75$th percentiles.
    Bottom: Maximum success probability against $C_\text{Ising}$ for the instances from $\mathcal{I}_\text{BP}$.}
    \label{fig:successVsJintraCoeffRatio}
\end{figure}

We used all embeddable instances from $\mathcal{I}_\text{BP}$ for investigating the performance of the D-Wave 2000Q.
As penalty weights, we used $\lambda_\text{one} = f^{\text{one}} T^{\text{one}}$ and $\lambda_\text{not} = f^{\text{not}} T^{\text{not}}$, with  $f^{\text{one}}, f^{\text{not}} \in \{\frac{1}{2}, 1\}$,
if the corresponding exact solution was valid.
Note, that this is always the case for the worst case estimation $f^{\text{one}} = f^{\text{not}} = 1$.
Again, we used 5 different embeddings for each problem instance.
The annealing solutions were obtained using $1000$ annealing runs, no gauging and majority voting as a un-embedding strategy for broken chains of logical qubits.
In order to calculate the time-to-solution with $99\%$ certainty we use
\begin{equation*}
	T_{99} = \frac{\ln(1 - 0.99)}{\ln(1-p)} T_{\text{anneal}},
\end{equation*}
where $T_{\text{anneal}}$ is the annealing time, which we fixed to $20\mu s$, and $p$ is the success probability.
The latter is calculated by the ratio of the number of runs where the optimal solution was found to the total number of runs.   
The best annealing solution was compared to an exact solution obtained with a MaxSAT solver \cite{kuegel2010}.
As intra-logical qubit coupling we used $J_F=-1$ in units of the largest coefficient of the embedded Ising model of the problem instance at hand.
The top part of figure~\ref{fig:successVsJintraCoeffRatio} shows the dependence of the success probability on the choice of $J_F$ for a single instance from $\mathcal{I}_\text{BP}$.
As expected,  for large $J_F$ the success probability is suppressed due to increased precision requirements and for very small $J_F$ the logical qubit chains are broken (cf.~\cite{stollenwerk2017}).
The former is substantiated by the bottom part of figure~\ref{fig:successVsJintraCoeffRatio}, where the success probability in dependence of $C_\text{Ising}$ is shown for a fixed $J_F=-1$ in units of the largest coefficient of the embedded Ising model.
The success probability is suppressed for larger values of $C_\text{Ising}$ due to the increased precision requirements.

\begin{figure}[t]
    \centering
    \includegraphics[width=0.9\linewidth]{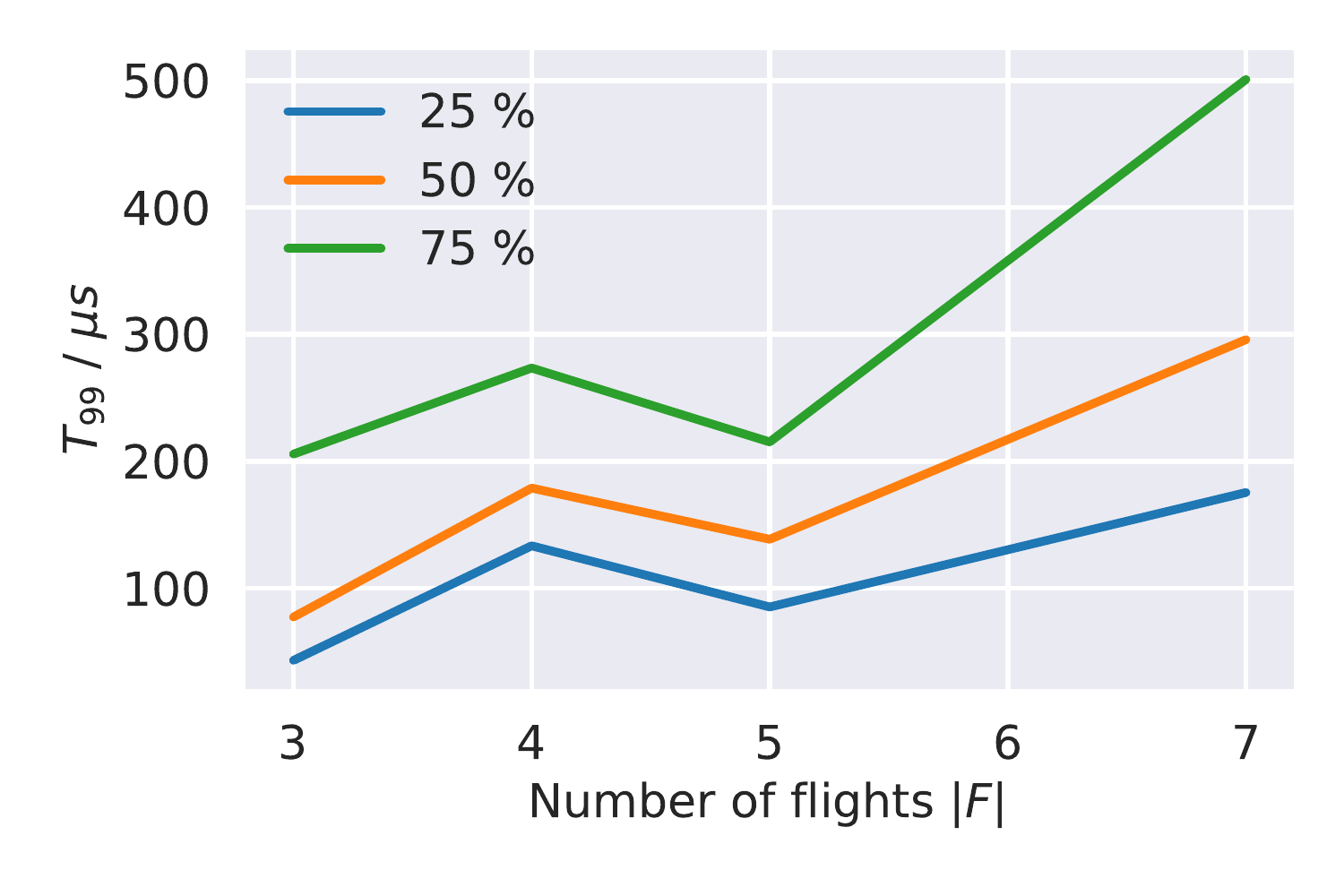}
    \includegraphics[width=0.9\linewidth]{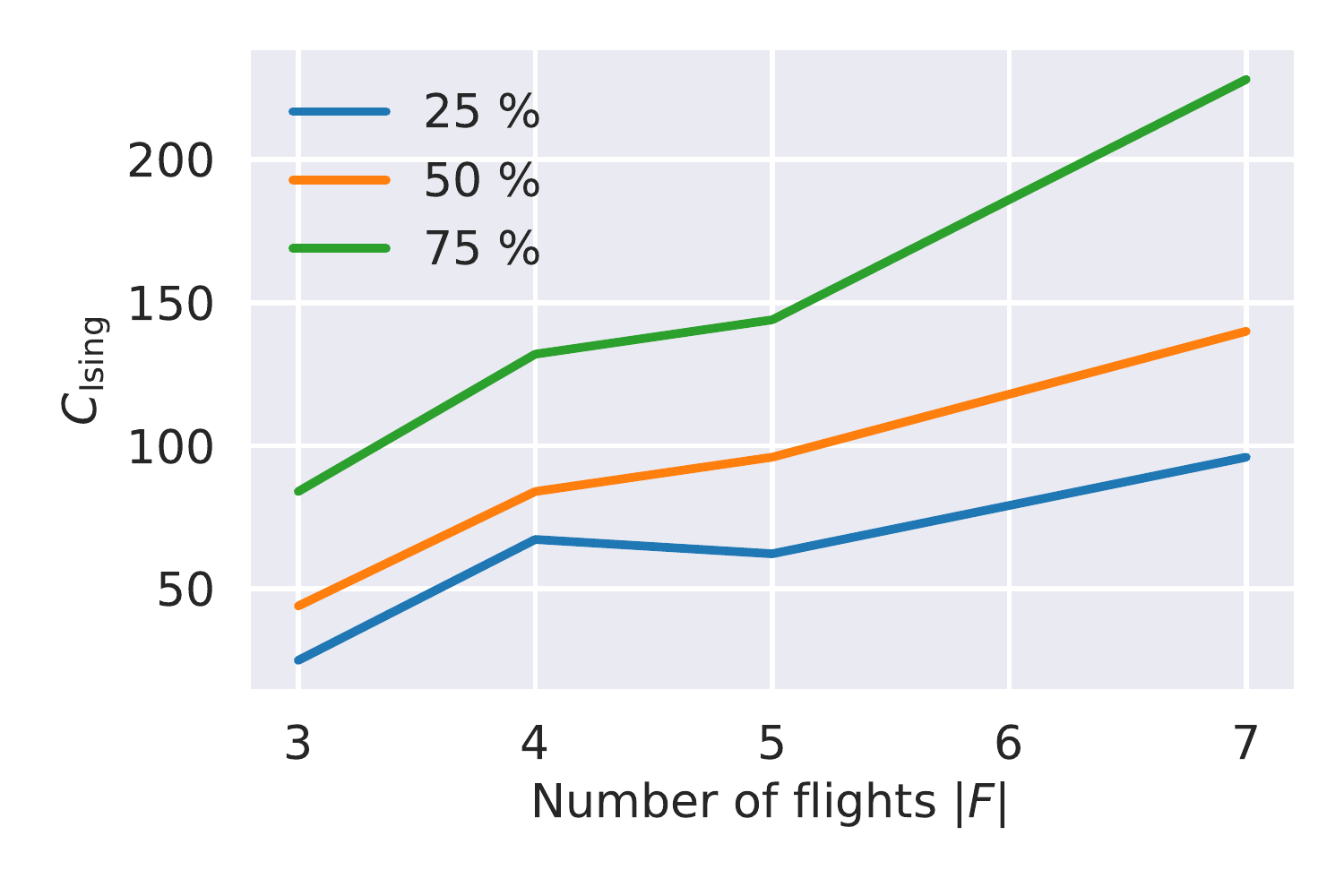}
    \caption{Top: Success probability against the number of flights for the instances from $\mathcal{I}_\text{BP}$.  
    Bottom: Maximum $C_\text{Ising}$ against the number of flights for the instances from $\mathcal{I}_\text{BP}$.
    The different colors represent the $25$th, $50$th and $75$th percentiles.}
    \label{fig:runtimeVsNumFlights}
\end{figure}
The top part of figure~\ref{fig:runtimeVsNumFlights} shows the time-to-solution in dependence of the number of flights.
There is an increase in the time-to-solution with the number of flights, and therefore the problem size.
This can be explained by the increase in $C_\text{Ising}$ with the number of flights as it can be seen on the bottom part of figure~\ref{fig:runtimeVsNumFlights}.

\section{Conclusion}

We showed, that the flight gate assignment problem can be solved with a quantum annealer with some restrictions.
First, the size of the amenable problems is very small.
Due to the high connectivity of the problem there is a large embedding overhead.
Therefore a conclusive assessment of the scaling behavior is not possible at the moment. 
Future generations of quantum annealers with more qubits and higher connectivity are needed to investigate larger problems.
Second, extracting problem instances directly from the data can lead to distributed coefficients in the resulting QUBOs.
As a result, the success probability is mostly suppressed for these instances due to their high precision requirements.
However, bin packing the coefficients can strongly decrease the precision requirements while retaining the heart of the problem.

For future work we leave the investigation of hybrid algorithms in order to recombine partial solutions and solve the whole problem, like in \cite{Tran},  
as well as the further investigation of the influence of bin packing on the solution quality.

\section{Acknowledgments}		

The authors would like to thank NASA Ames Quantum Artificial Intelligence Laboratory for their support during performing the experiments on the D-Wave 2000Q system,
for many valuable discussions and the opportunity to use the D-Wave 2000Q machine at NASA Ames.